\documentclass{article}
\usepackage{amssymb}

\usepackage{cite}
\usepackage{graphicx}
\usepackage{dcolumn}

\begin{document}

\title{On the straigtforward perturbation theory in classical mechanics}
\author{Paolo Amore \thanks{%
paolo.amore@gmail.com} \\
Facultad de Ciencias, Universidad de Colima, \\
Bernal D\'{\i}az del Castillo 340, Colima, Colima, Mexico\\
Francisco M. Fern\'andez \thanks{%
framfer@gmail.com}\\
INIFTA (CONICET), Divisi\'{o}n Qu\'{i}mica Te\'{o}rica,\\
Blvd. 113 y 64 (S/N), Sucursal 4, Casilla de Correo 16,\\
1900 La Plata, Argentina}
\maketitle

\begin{abstract}
We show that it is possible to extract useful information from the
straightforward perturbation theory in classical mechanics. Although the
secular terms make the perturbation series useless for large time, these
expansions yield the perturbation corrections for the period exactly and may
even be useful for sufficiently short time. We think that present analysis
may be suitable for an advanced undergraduate course on classical mechanics.
\end{abstract}

\section{Introduction}

\label{sec:intro}

Straightforward application of perturbation theory to classical periodic
systems gives rise to the so-called secular terms that are unbounded
perturbation corrections that spoil the periodic behaviour at sufficiently
large times. Such unwanted terms come from resonant contributions to the
perturbation equations. Most textbooks on classical mechanics discuss this
problem in detail and describe alternative perturbation approaches free from
secular termsl\cite{M65,N81}. One of them is the Lindstedt-Poincar\'{e}
method that consists on explicitly introducing the oscillation frequency
into the equations motion and setting its perturbation coefficients to
remove the resonant terms. In this way one obtains the perturbation
corrections to the trajectory and to the frequency at the same time\cite
{M65,N81}.

The purpose of this paper is to investigate the possibility of extracting
useful information from the perturbation series with secular terms. We
believe that the analysis of this old problem from another perspective may
be beneficial for undergraduate students that take an introductory course in
classical mechanics.

In section~\ref{sec:model} we introduce a simple model and convert
the equation of motion into a dimensionless differential equation.
In this way we reduce the number of independent model parameters.
In section\ref{sec:PT} we derive the straightforward perturbation
series with secular terms from the corresponding equations in
matrix form. In section\ref{sec:Lindstedt-Poincare} we outline the
lindstedt-Poincar\'{e} method and obtain the perturbation series
free from the unwanted secular terms (again using the matrix
approach). In section\ref{sec:comparison} we compare the results
of both series and show that the perturbation expansion with
secular terms provides some useful information about the periodic
system. Finally, in section\ref{sec:conclusions} we comment on the
main results of the paper and draw conclusions.

\section{The model}

\label{sec:model}

For simplicity and concreteness we consider a particle of mass $m$ under the
effect of the quartic potential
\begin{equation}
V(x)=\frac{k}{2}x^{2}+\frac{k^{\prime }}{4}x^{4}.
\end{equation}
The equation of motion is
\begin{equation}
m\ddot{x}(t)=-kx(t)-k^{\prime }x(t)^{3},  \label{eq:eq_mot}
\end{equation}
where a point indicates differentiation with respect to time, and we assume
the initial conditions $x(0)=x_{0}$, $\dot{x}(0)=0$.

If we define
\begin{equation}
\tau =\Omega _{0}t,\;\Omega _{0}=\sqrt{\frac{k}{m},\;}q=\frac{x}{x_{0}},
\end{equation}
the equation of motion (\ref{eq:eq_mot}) can be rewritten as
\begin{equation}
q^{\prime \prime }(\tau )=-q(\tau )-\lambda q(\tau )^{3},\;\lambda =\frac{%
k^{\prime }}{k}x_{0}^{2},  \label{eq:eq_mot_dim}
\end{equation}
where a prime indicates differentiation with respect to the dimensionless
time $\tau $ and $\lambda $ is a dimensionless perturbation parameter. The
initial conditions become $q(0)=1$ and $q^{\prime }(0)=0$.

\section{Straightforward perturbation theory}

\label{sec:PT}

In order to apply perturbation theory we rewrite the dimensionless equation
of motion (\ref{eq:eq_mot_dim}) as
\begin{eqnarray}
q^{\prime } &=&v,  \nonumber \\
v^{\prime } &=&-q-\lambda q^{3}.  \label{eq:eq_mot_dim_first_order}
\end{eqnarray}
If we expand
\begin{eqnarray}
q(\tau ) &=&\sum_{j=0}^{\infty }q_{j}(\tau )\lambda ^{j},  \nonumber \\
v(\tau ) &=&\sum_{j=0}^{\infty }v_{j}(\tau )\lambda ^{j},
\end{eqnarray}
we obtain the perturbation equations
\begin{eqnarray}
q_{n}^{\prime } &=&v_{n},  \nonumber \\
v_{n}^{\prime } &=&-q_{n}-\left( q^{3}\right) _{n-1},\;\left( q^{3}\right)
_{n-1}=\sum_{j=0}^{n-1}q_{n-1-j}\sum_{k=0}^{j}q_{k}q_{j-k}.
\label{eq:pert_eqs}
\end{eqnarray}
I is convenient to rewrite these equations in matrix form
\begin{eqnarray}
\mathbf{X}_{n}^{\prime } &=&\mathbf{KX}_{n}+\mathbf{R}_{n},  \nonumber \\
\mathbf{X}_{n} &=&\left(
\begin{array}{l}
x_{n} \\
v_{n}
\end{array}
\right) ,\;\mathbf{K}=\left(
\begin{array}{ll}
0 & 1 \\
-1 & 0
\end{array}
\right) ,\;\mathbf{R}_{n}=\left(
\begin{array}{l}
0 \\
-\left( q^{3}\right) _{n-1}
\end{array}
\right) ,  \label{eq:pert_eqs_matrix}
\end{eqnarray}
because its solution is straightforward
\begin{equation}
\mathbf{X}_{n}(\tau )=\exp \left( \mathbf{K}\tau \right) \left[ \mathbf{W}%
_{n}+\int_{0}^{\tau }\exp \left( -\mathbf{K}s\right) \mathbf{R}_{n}\right]
,\;\mathbf{W}_{n}=\left(
\begin{array}{l}
a_{n} \\
b_{n}
\end{array}
\right) .  \label{eq:PT_sol_matrix}
\end{equation}
The analytical calculation of
\begin{equation}
\exp \left( \mathbf{K}\tau \right) =\left(
\begin{array}{ll}
\cos (\tau ) & \sin (\tau ) \\
-\sin (\tau ) & \cos (\tau )
\end{array}
\right) ,
\end{equation}
offers no difficulty\cite{A69,FC96}. At order zero we have
\begin{equation}
q_{0}(\tau )=\cos (\tau ),\;v_{0}=-\sin (\tau ),
\end{equation}
consistent with the initial conditions $q(0)=1$ and $v(0)=0$ provided that $%
a_{n}=b_{n}=0$ for all $n>0$.

The first two perturbation corrections for the dimensionless coordinate $%
q(\tau )$ and velocity $v(\tau )$%
\begin{eqnarray}
q_{1}(\tau ) &=&\frac{\cos {\left( 3\tau \right) }}{32}-\frac{\cos {\left(
\tau \right) }}{32}-\frac{3\tau \sin {\left( \tau \right) }}{8},  \nonumber
\\
q_{2}(\tau ) &=&\frac{\cos {\left( 5\tau \right) }}{1024}-\frac{3\cos {%
\left( 3\tau \right) }}{128}+\frac{23\cos {\left( \tau \right) }}{1024}+\tau
\left[ \frac{3\sin {\left( \tau \right) }}{32}-\frac{9\sin {\left( 3\tau
\right) }}{256}\right]   \nonumber \\
&&-\frac{9\tau ^{2}\cos {\left( \tau \right) }}{128},  \nonumber \\
v_{1}(\tau ) &=&-\frac{11\sin {\left( \tau \right) }}{32}-\frac{3\sin {%
\left( 3\tau \right) }}{32}-\frac{3\tau \cos {\left( \tau \right) }}{8},
\nonumber \\
v_{2}(\tau ) &=&\frac{73\sin {\left( \tau \right) }}{1024}+\frac{9\sin {%
\left( 3\tau \right) }}{256}-\frac{5\sin {\left( 5\tau \right) }}{1024}-%
\frac{3\tau \cos {\left( \tau \right) }}{64}-\frac{27\tau \cos {\left( 3\tau
\right) }}{256}  \nonumber \\
&&+\frac{9\tau ^{2}\sin {\left( \tau \right) }}{128},  \label{eq:q1,q2}
\end{eqnarray}
exhibit secular terms proportional to powers of $\tau $ that are unbounded
and in principle inconsistent with a periodic motion\cite{M65,N81}.

\section{The Lindstedt-Poincar\'{e} method}

\label{sec:Lindstedt-Poincare}

One of the standard strategies for removing the secular terms from the
straightforward perturbation theory is the Lindstedt-Poincar\'{e} method\cite
{M65,N81} that consists of defining a new time variable $s=\omega \tau $,
where $\omega $ is the dimensionless frequency of the periodic motion (if $%
\Omega $ is the actual frequency then $\omega =\Omega /\Omega _{0}$). In
this way the equations of motion (\ref{eq:eq_mot_dim_first_order}) become
\begin{eqnarray}
\omega \xi ^{\prime }(s) &=&\eta (s),  \nonumber \\
\omega \eta ^{\prime }(s) &=&-\xi (s)-\lambda \xi (s)^{3},
\end{eqnarray}
where $\xi (s)=q(s/\omega )$ and $\eta (s)=v(s/\omega )$.

If we expand
\begin{eqnarray}
\xi (s) &=&\sum_{j=0}^{\infty }\xi _{j}(s)\lambda ^{j},  \nonumber \\
\eta (s) &=&\sum_{j=0}^{\infty }\eta _{j}(s)\lambda ^{j},  \nonumber \\
\omega (s) &=&\sum_{j=0}^{\infty }\omega _{j}(s)\lambda ^{j},\;\omega _{0}=1,
\end{eqnarray}
the perturbation equations become
\begin{eqnarray}
\xi _{n}^{\prime } &=&\eta _{n}-\sum_{j=1}^{n}\omega _{j}(s)\xi
_{n-j}^{\prime },  \nonumber \\
\eta _{n}^{\prime } &=&-\xi _{n}-\left( \xi ^{3}\right)
_{n-1}-\sum_{j=1}^{n}\omega _{j}(s)\eta _{n-j}^{\prime }.
\end{eqnarray}
We can solve these equations very easily in matrix form by just modifying
the column vector $\mathbf{R}_{n}$ in equation (\ref{eq:PT_sol_matrix})
because the matrix $\mathbf{K}$ is exactly the same.

The resulting equations depend on the expansion coefficients $\omega _{j}$
that we choose so that the secular terms vanish\cite{M65,N81}. In this way
we obtain
\begin{eqnarray}
\xi _{1}(s) &=&-\frac{1}{32}\left[ \cos {\left( s\right) }+\cos {\left(
3s\right) }\right] ,  \nonumber \\
\eta _{1}(s) &=&-\frac{1}{32}\left[ 11\sin {\left( s\right) }+3\sin {\left(
3s\right) }\right] ,  \nonumber \\
\xi _{2}(s) &=&\frac{1}{1024}\left[ 23\cos {\left( s\right) }-24\cos {\left(
3s\right) }+\cos {\left( 5s\right) }\right] ,  \nonumber \\
\eta _{2}(s) &=&\frac{1}{1024}\left[ 73\sin {\left( s\right) }+36\sin {%
\left( 3s\right) }-5\sin {\left( 5s\right) }\right] ,  \label{eq:xi_1,xi_2}
\end{eqnarray}
that are periodic because the perturbation coefficients for the frequency
\begin{eqnarray}
\omega _{1} &=&\frac{3}{8},  \nonumber \\
\omega _{2} &=&-\frac{21}{256}.  \label{eq:omega_1,omega_2}
\end{eqnarray}
where chosen in such a way that the coefficients of the secular terms
vanish. We do not enlarge upon this particular point because it is discussed
in detail in most textbooks\cite{M65,N81}.

Equations (\ref{eq:xi_1,xi_2}) suggest that
\begin{eqnarray}
\xi _{j}(s) &=&\sum_{k=0}^{j}a_{kj}\cos \left[ (2k+1)s\right] ,  \nonumber \\
\eta _{j}(s) &=&\sum_{k=0}^{j}b_{kj}\sin \left[ (2k+1)s\right] .
\end{eqnarray}
For example, the next two terms are
\begin{eqnarray}
a_{03} &=&-\frac{547}{32768},\;a_{13}=\frac{297}{16384},\;a_{23}=-\frac{3}{%
2048},\;a_{33}=\frac{1}{32768},  \nonumber \\
b_{03} &=&-\frac{1109}{32768},\;b_{13}=-\frac{333}{16384},\;b_{23}=\frac{45}{%
8192},\;b_{33}=-\frac{7}{32768},  \nonumber \\
a_{04} &=&\frac{6713}{524288},\;a_{14}=-\frac{15121}{1048576},\;a_{24}=\frac{%
883}{524288},\;a_{34}=-\frac{9}{131072},\;  \nonumber \\
a_{44} &=&\frac{1}{1048576},  \nonumber \\
b_{04} &=&\frac{11281}{524288},\;b_{14}=\frac{14043}{1048576},\;b_{24}=-%
\frac{2765}{524288},\;b_{34}=\frac{105}{262144},\;  \nonumber \\
b_{44} &=&-\frac{9}{1048576},
\end{eqnarray}
provided that
\begin{eqnarray}
\omega _{3} &=&\frac{81}{2048},  \nonumber \\
\omega _{4} &=&-\frac{6549}{262144}.  \label{eq:omega3,omega4}
\end{eqnarray}
In order to compare equations (\ref{eq:q1,q2}) and (\ref{eq:xi_1,xi_2}) we
should take into account that $s=\omega \tau $, where $\omega \approx
1+\omega _{1}\lambda +\omega _{2}\lambda ^{2}$ at second order. Clearly, the
period of the perturbation corrections (\ref{eq:xi_1,xi_2}) in terms of the
dimensionless time $\tau $ is $T=2\pi /\omega $.

\section{Comparison of the perturbation results}

\label{sec:comparison}

In order to compare the perturbation approaches with and without the secular
terms we define the partial sums
\begin{equation}
q^{[N]}(\tau )=\sum_{j=0}^{N}q_{j}(\tau )\lambda ^{j},
\end{equation}
and similar expressions for $v$, $\xi $, $\eta $, etc.. We first try to
estimate the perturbation corrections to the dimensionless period $T$ from
the equation
\begin{eqnarray}
q^{[N]}\left( T^{[N]}\right) -q^{[N]}(0) &=&\mathcal{O}\left( \lambda
^{N+1}\right) ,  \nonumber \\
T^{[N]} &=&\sum_{j=0}^{N}T_{j}\lambda ^{j},\;T_{0}=2\pi .
\end{eqnarray}
The calculation exhibits some surprising features; for example
\begin{equation}
q^{[3]}\left( T^{[1]}\right) -q^{[3]}(0)=\mathcal{O}\left( \lambda
^{4}\right) ,
\end{equation}
with just
\begin{equation}
T_{1}=-\frac{3\pi }{4}.  \label{eq:T_1}
\end{equation}
On the other hand, from
\begin{equation}
v^{[4]}\left( T^{[4]}\right) -v^{[4]}(0)=\mathcal{O}\left( \lambda
^{5}\right) ,
\end{equation}
we obtain (\ref{eq:T_1}) as well as
\begin{eqnarray}
T_{2} &=&\frac{57\pi }{128},  \nonumber \\
T_{3} &=&-\frac{315\pi }{1024},  \nonumber \\
T_{4} &=&\frac{30345\pi }{131072}.  \label{eq:T_2,T_3,T_4}
\end{eqnarray}
We conclude that it is more convenient to obtain the expansion for the
period from $v^{[N]}(\tau )$ although $q^{[N]}(\tau )$ appears to be more
accurate for this model. In any case we appreciate that the unwanted secular
terms in the perturbation series do not prevent us from obtaining the
correct expansion for the period. Note that the perturbation corrections $%
\omega _{j}$ obtained above by means of the Lindstedt-Poincar\'{e} method
are related to the perturbation corrections $T_{k}$ obtained here by
\begin{equation}
\frac{2\pi }{\omega ^{[N]}}=T^{[N]}+\mathcal{O}\left( \lambda ^{N+1}\right) .
\end{equation}
All these results appear to suggest that the perturbation series with
secular terms may be reasonably accurate in the interval $0\leq \tau
\lesssim T$.

\begin{figure}[tbp]
\begin{center}
\includegraphics[width=9cm]{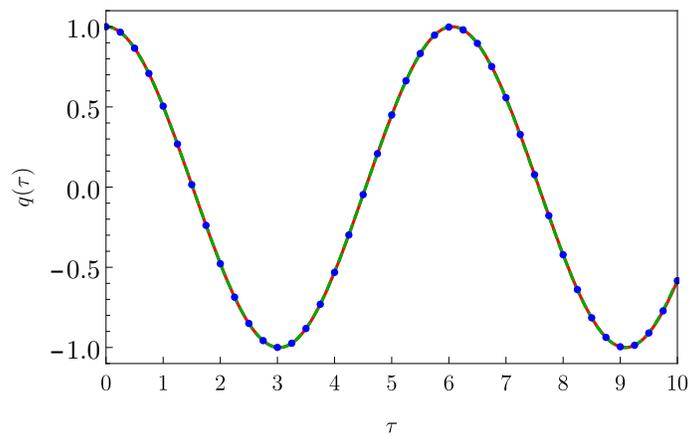}
\par
\end{center}
\caption{Exact $q(\tau)$ (solid circles, blue), perturbation theory with
secular terms (dashed line, green) and Lindstedt-Poincar\'e series (solid
line, red) for $\lambda=0.1$}
\label{Fig:QT01}
\end{figure}

In order to verify the conclusions drawn above we calculated $q^{[2]}(\tau )$
with both perturbation approaches and compared them with the exact result
(obtained numerically with sufficient precision) for moderate times (no much
larger than $T$). Fig.~\ref{Fig:QT01} shows that the agreement is remarkable
for $\lambda =0.1$, whereas Fig.~\ref{fig:q,v,l05} shows that $q^{[2]}$ is
noticeably more accurate than $v^{[2]}$ when $\lambda =0.5$. In the latter
case the Lindstedt-Poincar\'{e} series to identical order are considerably
more accurate than the straightforward expansions as expected.

\section{Conclusions}

\label{sec:conclusions}

It is clear that the perturbation series with secular terms are unsuitable
for the description of the periodic motion of the oscillator for
sufficiently large $\tau $ and for this reason one commonly resorts to the
Lindstedt-Poincar\'{e} expansion. This well known fact is discussed in most
textbooks on classical mechanics\cite{M65,N81}. However, the former series
provide considerable information about the dynamics of the oscillator. We
have shown that they yield the exact perturbation expansion for the period
and, if $\lambda $ is not too large, they also give a reasonable estimate of
the motion when $0\leq \tau \leq T$. In such a case we can use these results
for all $\tau $ because we know that the exact coordinate satisfies $q(\tau
+T)=q(\tau )$.

We think that present discussion of the perturbation series for a simple
model may be of interest in an undergraduate course on classical mechanics.
Since the calculation of perturbation corrections of order larger than the
second one by hand may be extremely tedious and error prone, present
pedagogical proposal may suitable for stimulating the application of
available computer algebra systems. Reasonable mastering of such software is
certainly beneficial for physics students because it enables them trying
somewhat complicated analytical calculations without spending too much time.
The matrix method outlined in this paper is particular useful for the
application of computer algebra systems.

\begin{figure}[tbp]
\begin{center}
\includegraphics[width=9cm]{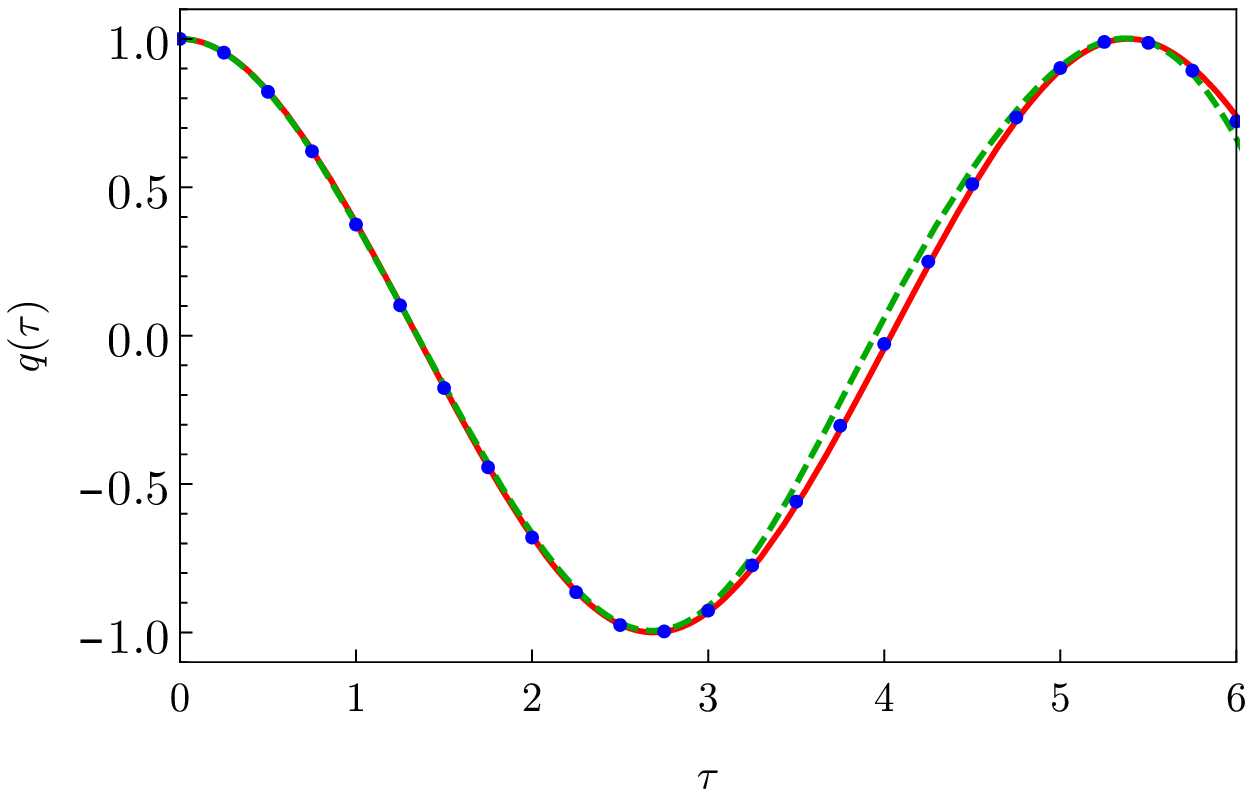} \includegraphics[width=9cm]{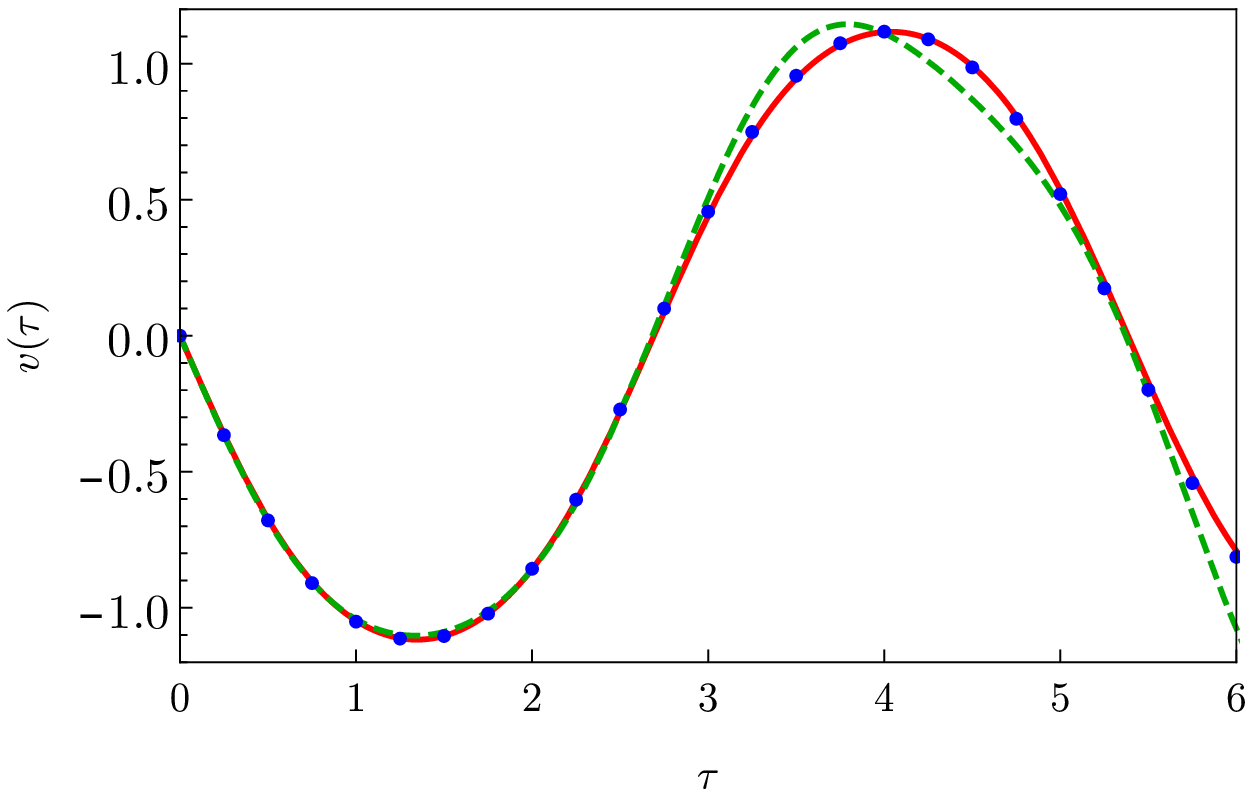}
\par
\end{center}
\caption{Exact $q(\tau)$ (solid circles, blue), perturbation theory with
secular terms (dashed line, green) and Lindstedt-Poincar\'e series (solid
line, red) for $\lambda=0.5$ and $v(\tau)$ (lower figure) for $\lambda=0.5$}
\label{fig:q,v,l05}
\end{figure}

\end{document}